\documentclass[preprint,12pt]{elsarticle}

 \usepackage{graphics}
 \usepackage{epsfig}

\usepackage{amssymb}
\newcommand\nn{\nonumber}
\newcommand\ba{\begin{eqnarray}}
\newcommand\ea{\end{eqnarray}}
\newcommand\be{\begin{equation}}
\newcommand\ee{\end{equation}}

\journal{Nuclear Physics A}

\begin{document}

\begin{frontmatter}


 \title{Model independent study of massive lepton elastic scattering on the proton, beyond the Born approximation}
\author{G. I. Gakh}
% \ead{gakh@kipt.kharkov.ua}
 \address{National Science Centre, Kharkov Institute of Physics and Technology, \\ 61108 Akademicheskaya 1, Kharkov,
Ukraine and \\ 
V. N.
Karazin Kharkov National University, Dept. of Physics and
Technology, \\ 31 Kurchatov, 61108, Kharkov, Ukraine}
\author{M. Konchatnyi}
\address{National Science Centre, Kharkov Institute of Physics and Technology, \\ 61108 Akademicheskaya 1, Kharkov,
Ukraine}
\author{A. Dbeyssi} 
\address {Helmholtz Institute Mainz, Johann-Joachim-Becher-Weg 45, D-55099 Mainz,
Germany}
\author{E. Tomasi--Gustafsson\corref{cor1}\fnref{label2}}
%\author{ E. Tomasi--Gustafsson\corref{cor1}\fnref{label2}}
\cortext[cor1]{Corresponding author: E. Tomasi-Gustafsson (egle.tomasi@cea.fr)}
\fntext[label2]{\it Permanent address: }
\address{\it DSM/IRFU/SPhN, CEA/Saclay, 91191 Gif-sur-Yvette  and \\Univ Paris-Sud, CNRS/IN2P3, Institut de Physique 
Nucl\'eaire, UMR 8608, \\ 91406 Orsay, France}
 
%%%%%%%%%%%%%%%%%%%%%%%%%%%%%%%%%%%%%%
\begin{abstract}
%%%%%%%%%%%%%%%%%%%%%%%%%%%%%%%%%%%%%%
Model independent expressions for all polarization observables in $\mu+p\to \mu+p$ elastic scattering are obtained, taking into account the lepton mass 
and including the two-photon exchange contribution.
The spin structure of the matrix element is parametrized in terms of six
independent complex amplitudes, functions of two independent kinematical variables. 
General statements about the influence of the
two--photon--exchange terms on the differential cross section and on 
polarization observables are given. Polarization effects
have been investigated for the case of a longitudinally polarized
lepton beam and polarized nucleon in
the  final state.
\end{abstract}


\end{frontmatter}

% \linenumbers

\section{Introduction}
\label{Introduction}

%%%%%%%%%%%%%%%%%%%%%%%%%%%%%%%%%%%%%%

In this work, elastic lepton nucleon scattering is revised, deriving
the kinematical relations and the expressions of the observables in
the most general form, taking into account the lepton mass and
expressing the matrix element as function of six independent
amplitudes.

Electron proton elastic scattering, which involves four spin
1/2 particles, is considered one of the simplest
reaction to investigate the proton structure.  
Assuming that the interaction occurs through the
exchange of a virtual photon, considering the spin one nature of the
virtual photon, parity conservation, and the identity of the initial
and final state bring the number of amplitudes from sixteen to two.
These two amplitudes, in general complex,  become
real due to unitarity as the momentum transfer $q^2=-Q^2$ is negative (space-like region).

The cross section is expressed as function of two form factors (FFs),  electric $G_E$ and magnetic $G_M$, as given in Ref.
\cite{Rosenbluth:1950yq}. This formula assumes that the interaction
occurs through the exchange of a virtual photon, which carries the
transfer momentum $q$ (one photon exchange approximation) and
neglects the electron mass, which are considered good approximations
at energies of the order of few GeV. Polarization observables  for elastic electron proton scattering were
derived in \cite{Akhiezer:1968ek,Akhiezer:1974em}.

Several experiments were carried out, starting from the ones for
which Hofstadter was rewarded by Nobel prize in 1966, to the recent
polarization measurements from the GEp collaboration at JLab \cite
{Puckett:2010ac}. The surprising result was that polarized and unpolarized
experiments, although based on the same theoretical background (same formalism and same
assumptions), ended up with inconsistent values of the FF ratio. Note
that in the unpolarized case, one measures the cross section at 
fixed $Q^2$ $(Q^2=-q^2>0) $, changing the beam energy and the scattered electron angle, 
which gives access to $G_E^2$ and $G_M^2$, whereas, in
the second case, one measures the ratio of the longitudinal to transverse
recoil proton polarization which is directly related to the ratio of $G_E$ and $G_M$. 

It turned out that ratio of the electric to the magnetic FF, reconstructed from
unpolarized measurements, is close to unity and constant with $Q^2$ 
whereas the polarization method gives a monotonically decreasing
behavior, and a value close to zero at $Q^2\le 9$ GeV$^2$.
This result, as well as an eventual zero crossing of this ratio, have given rise to
a large theoretical and experimental work. It was suggested
that the reaction mechanism (one-photon-exchange) is not a good
approximation at large $Q^2$ \cite{Guichon:2003qm}. The idea that
two-photon exchange could play a role at large momentum transfer,
was firstly discussed in the 70's \cite{Boitsov:1972if,Franco:1973uq, Gunion:1972bj}, and more recently, for electron
deuteron scattering (where a discrepancy between two sets of data
was also found in measurements of the structure function $A(Q^2)$
\cite{Rekalo:1999mt}). 

In presence of two photon exchange, the number 
of amplitudes which describe the reaction is  
essentially increased. Instead of two real amplitudes, function of one
variable, $Q^2$, one has three complex amplitudes (in the approximation of zero lepton mass), functions of two
variables, for example the electron scattering angle and $Q^2$. 
In Refs. \cite{Rekalo:2003xa,Rekalo:2003km,Rekalo:2004wa} it was shown, 
that, in such conditions, one can still
extract  the real FFs, $G_E$ and $G_M$, but for this aim, it is
necessary to measure three T-odd or five T-even polarization
observables, including triple spin observables which are expected to
be of the order of the electromagnetic coupling constant, $\alpha$.
Alternatively, one can use polarized electron and positron beams, in the same
kinematical conditions and apply a generalized polarization method.

Moreover, some of the polarization observables are proportional to
the lepton mass.  In this case, not only helicity conserving amplitudes, but
also spin--flip amplitudes should be taken into account, as they are of the same order of magnitude. The
number of relevant amplitudes from three becomes six. The additional 
amplitudes have been neglected in most of the works related to two
photon exchange. However, if one looks to observables which are
expected to be small, or to heavy lepton scattering, these
amplitudes should be taken into account.

Muon-proton scattering brings the same information concerning the electromagnetic FFs. However, in case of low energy and large lepton mass the terms
proportional to the lepton mass become important and the mass should
be taken explicitly into account. The expressions of the kinematical
relations and of the polarized and unpolarized observables are
different from those currently used. In Ref. \cite{Preedom:1987mx} the 
effect of the lepton mass was discussed for $ep$ elastic scattering.
The unpolarized cross section and the double 
spin asymmetry, when the lepton beam and the target are polarized,
were calculated, in the one-photon approximation. 

The terms related to the lepton mass influence also the extraction of the charge radius of the proton
from muon proton elastic scattering,
which is related, in non relativistic approximation, to the limit at
$Q^2\to 0$ of the derivative of the electric FF. Note that the
problem of the proton size is object of large interest, due to the
recent experiment on muonic hydrogen by laser spectroscopy
measurement of the $\nu p$(2S-2P) transition frequency
\cite{Pohl:2010zza}. The result on the proton charge radius
$r_c=0.84184(67)$ fm obtained in this experiment is one order of
magnitude more precise but smaller by five standard deviation
compared to the best value previously assumed  $r_c=0.8768(69)$ fm (CODATA \cite{Mohr:2008fa}). 
Previous best measurements include
techniques based on hydrogen spectroscopy, which are more precise, but
compatible with electron proton elastic scattering at small values
of the four momentum transfer squared $Q^2$. The most recent result
from electron proton elastic scattering, $r_c=0.879(5)_{\rm
stat}(4)_{\rm syst}(2)_{\rm model}(4)_{\rm group}$ fm, can be found
in Ref. \cite{Bernauer:2010wm}. The smallest value of $Q^2$ reached
with good precision was of the order of 10$^{-3}$ GeV$^2$. 

In Ref.  \cite{Gakh:2011sh} the elastic scattering of a proton beam
on an electron target has been considered. In
this case it is necessary to take into account the lepton mass, due to the inverse kinematics (the
projectile is heavier than the target).
This reaction allows to access the region of very small transferred momenta, up to
$10^{-6}$ GeV$^2$. 

The fact that the proton charge radius was not measured in the
process of the elastic muon-proton scattering led to the proposal 
{\it  MUon proton Scattering Experiment} (MUSE) at Paul Scherrer
Institut \cite{Gilman:2013eiv}. 

This experiment plans 
a simultaneous measurement of the elastic $\mu^-p$ and $e^-p$ scattering  as well as $\mu^+ p$ and $e^+p$. It will establish the consistency or the 
difference of the muon-proton and electron-proton interaction with good precision in the considered kinematics \cite{Gilman:2013eiv}. Three values of the muon beam momenta which are comparable with the muon mass: 115 MeV, 153 MeV, and 210
MeV, were chosen.  However, in case of low
energy and large lepton mass, the terms proportional to the lepton
mass become important and the mass should be taken explicitly into
account in the calculation of the kinematical variables and of the experimental observables. The MUSE experiment intends to observe, if any, two-photon-exchange effects and to test their effect on the radius
extraction. The relative precision of the cross section measurements  will be 
about few tenth of a percent.

The muon-proton interaction is expected to have
reduced radiative corrections compared to electrons, which gives a
safer extraction of the physics information related to the proton
structure. Moreover, the muon beam being naturally polarized, 
when it is formed from pion decay, double spin polarization experiments 
can be foreseen, requiring to polarize the hydrogen target, 
or to measure the polarization of the outgoing
proton. The expressions for the polarization observables are also
affected by the lepton mass.

In this paper a complete derivation of polarized and
unpolarized lepton proton scattering, including the six amplitudes
(in presence of two-photon exchange) and taking into account the lepton mass, is presented. 

Experimental observables
are then derived in model independent way, as functions of the six
amplitudes. We  point out the relevant terms depending on the
lepton mass and on the new amplitudes, and show in which limits the
known formulas can be recovered,  for an appropriate use of the
approximated formulas existing in the literature.

%\section{Differential cross section}
%\hspace{0.7cm}
%%%%%%%%%%%%%%%%%%%%%%%%%%%%%%%%%
\section{Formalism}
%%%%%%%%%%%%%%%%%%%%%%%%%%%%%%%%%%
Let us consider the reaction:
\be
\mu(k_1)+p(p_1) \to \mu(k_2) +p(p_2),
\label{eq:eq1}
\ee
where the momenta of the particles are written in the parenthesis.
In the laboratory system, where the present analysis is performed, the proton
(muon) four momenta in the initial and final states are respectively 
$p_1$ and $p_2$ ($k_1$ and $k_2$) with components:
\be
p_1=(M,0),~ p_2=(E_2, \vec p_2),~ k_1=(\varepsilon_1, \vec k_1),~
k_2=(\varepsilon_2, \vec k_2),
\label{eq:eq2}
\ee
where $M$ is the proton mass. The general structure
of the matrix element of the reaction (\ref{eq:eq1}), taking into account the lepton mass and the two-photon-
exchange contribution, can be written as sum of two terms:
\be
{\cal M} ={\cal M}_1+{\cal M}_2,
\label{eq:eq3}
\ee
where the first (second) term is determined by the spin non-flip
(flip) amplitudes. The matrix element which is determined by the
spin non-flip amplitudes can be written as
\be
{\cal M}_1=\frac{e^2}{(-q^2)}j_{\mu}J_{\mu}, \ \ j_{\mu}=\bar
u(k_2)\gamma_{\mu }u(k_1),
\label{eq:eq4}
\ee
$$J_{\mu}=\bar u(p_2)\biggl [\tilde G_{M}(q^2, s)\gamma_{\mu
}-\frac{1}{M}\tilde F_{2}(q^2, s)P_{\mu}+\frac{1}{M^2}\tilde
F_{3}(q^2, s)P_{\mu}\hat K\biggr ]u(p_1), $$ 
where
$q=k_1-k_2=p_2-p_1,$, $s=(k_1+p_1)^2,$ $K=(k_1+k_2)/2$,
$P=(p_1+p_2)/2$. The second term in (\ref{eq:eq3}) is determined by the spin
flip amplitudes. Since these
amplitudes are proportional to the lepton mass $m$, we single out
explicitly the factor $m/M$ and the matrix element can be written as 
\ba
{\cal M}_2&=&\frac{m}{M}\frac{e^2}{Q^2}\biggl [\bar u(k_2)u(k_1)\bar
u(p_2)\biggl (\tilde F_{4}(Q^2, s)+\frac{1}{M}\tilde F_{5}(Q^2,
s)\hat K\biggr )u(p_1)+\nn\\
&&
\tilde F_{6}(Q^2, s)\bar
u(k_2)\gamma_{5}u(k_1)\bar u(p_2)\gamma_{5}u(k_p)\biggr ].
\label{eq:eq5}
\ea
where $m$ is the lepton mass. The six complex amplitudes, $\tilde G_{M}(q^2, s)$ and $\tilde
F_i(q^2, s), i=2-6, $ which are generally the functions of two
independent kinematical variables, $q^2$ and $s$, fully describe the
spin structure of the matrix element for the reaction (\ref{eq:eq1}), for any
number of changed virtual photons. 

This expression holds under assumption of the parity (P)--invariance of the
electromagnetic interaction. Note, however, that expressions (\ref{eq:eq4}) and
(\ref{eq:eq5}) are not unique, but many equivalent representations of the
$\mu^-p\rightarrow \mu^-p$ reaction matrix element may be written.

In the Born (one--photon--exchange) approximation these amplitudes reduce to:
\be
\tilde G_{M}^{Born}(q^2, s)=G_{M}(q^2), \  \tilde F_{2}^{Born}(q^2,
s)=F_{2}(q^2), \ \tilde F_i^{Born}(q^2, s)=0, \ i=3-6, \
\label{eq:eq6}
\ee
where $G_{M}(q^2)$ and $F_{2}(q^2)$ are the magnetic and Pauli
nucleon electromagnetic FFs, respectively, which are 
real functions of the variable $q^2$ in the space-like region
of the momentum transfer squared. In the following we use the
standard magnetic $G_{M}(q^2)$ and charge $G_{E}(q^2)$ nucleon form
factors, which are related to the  Pauli FF $F_{2}(q^2)$ by:
\be
F_{2}(q^2)=\frac{1}{1+\tau}(G_{M}(q^2)-G_{E}(q^2)), \ \tau =-\frac{q^2}{4M^2}.
\label{eq:eq7}
\ee

In analogy with this relation, let us introduce the linear
combinations of the $\tilde G_{M,E}(q^2, s)$ amplitudes which reduce to the Pauli FF $F_{2}$
in the Born approximation:
$$
\tilde F_{2}(q^2, s)=\frac{1}{1+\tau}(\tilde G_{M}(q^2, s)-\tilde G_{E}(q^2, s)).
$$
To separate the effects due to the Born and to the two-photon-exchange
contribution as well as the terms induced by the lepton mass, let us
single out the dominant contribution and define the following
decompositions of the amplitudes
\be
\tilde G_M(q^2, s)=G_M(q^2)+\Delta G_M(q^2, s), \ \tilde
G_E(q^2, s)=G_E(q^2)+\Delta G_E(q^2, s). \
\label{eq:eq8}
\ee
The order of magnitude of these quantities is 
$$\Delta G_M(q^2,s)\sim \Delta G_{E}(q^2, s)~
\sim \tilde F_i(q^2, s) (i=3-6)\sim
\alpha  ,~ G_{M,E}\sim \alpha^0 .$$
Since the terms $\Delta G_{M},
~\Delta G_{E}$ and $\tilde F_i$ are small in comparison with the
dominant ones, we neglect, in the following, the bilinear combinations
of these small terms.

The modulus of the matrix element squared in such
approximation can be written as
\be
|{\cal M}|^2 =|{\cal M}_1|^2+2Re{\cal M}_1{\cal M}_2^*.
\label{eq:eq9}
\ee
The differential cross section in terms of the modulus of the matrix element squared is
\be
d\sigma=\frac{(2\pi )^4}{4I}|{\cal M}|^2\frac{d{\vec k}_2d{\vec
p}_2}{(2\pi )^64\varepsilon_2E_2}\delta^{(4)}(k_1+p_1-k_2-p_2),
\label{eq:eq10}
\ee
where $I^2=(k_1\cdot p_1)^2-m^2M^2$ and $\varepsilon_2 (E_2)$ is the
energy of the scattered muon (recoil proton).

Writing the matrix element in the form 
${\cal M}=(e^2/(-q^2))\overline{\cal M}$, one can obtain following expression for the differential cross
section of the reaction (\ref{eq:eq1}) in the laboratory system for the case
when the scattered muon is detected in the final state
\be
\frac{d\sigma}{d\Omega}=\frac{\alpha ^2}{4M}\frac{{\vec
k}_2^2}{d|{\vec k}_1|}\frac{|\overline{\cal M}|^2}{q^4},
\label{eq:eq11}
\ee
where $d=(M+\varepsilon_1)|{\vec k}_2|-\varepsilon_2|{\vec
k}_1|\cos\theta  $, $\theta $  is the muon scattering angle (angle
between the directions of the initial and final muons), $d \Omega$ is the differential solid angle of the scattered muon). The scattered muon
energy is written in term of the muon scattering angle as
\be
\varepsilon_2=\frac{(\varepsilon_1+M)(M\varepsilon_1 +
m^2)+\vec{k}_1^2\cos\theta\sqrt{ M^2-m^2\sin^2\theta}}
{(\varepsilon_1+M)^2-\vec{k}_1^2\cos^2\theta }.
\label{eq:eq12}
\ee

The differential cross section for the case when the recoil proton
is detected in the final state ($d \Omega_p$ is the differential solid angle of the scattered proton) is:
\be
\frac{d\sigma}{d\Omega_p}=\frac{\alpha ^2}{4M}\frac{{\vec p}_2
^2}{\bar d|{\vec k}_1|}\frac{|\bar{\it M}|^2}{q^4},
\label{eq:eq13}
\ee
where $\bar d=(M+\varepsilon_1)|{\vec p}_2|-E_2|{\vec
k}_1|\cos\theta_p $,  and $\theta_p $  is the angle between the directions 
of the muon beam and the recoil proton. Using the relation
$$dq^2=-|{\vec k}_1||{\vec
p}_2|\frac{1}{\pi}\frac{E_2+M}{\varepsilon_1+M}d\Omega_p, $$ 
we obtain the following expression for the differential cross section over the $q^2$
variable
\be
\frac{d\sigma}{dq^2}=-\frac{\pi\alpha ^2}{4M}\frac{{|\vec p}_2|}{\bar
d{\vec k}_1^2}\frac{\varepsilon_1+M}{E_2+M}\frac{|\overline{\cal M}|^2}{q^4}.
\label{eq:eq14}
\ee
The modulus of the first matrix element squared can be written as
\be
|\overline{\cal M}_1|^2 =L_{\mu\nu }H_{\mu\nu },
\label{eq:eq15}
\ee
where the leptonic $L_{\mu\nu }$ and hadronic $H_{\mu\nu }$ tensors
are defined as follows
\be
L_{\mu\nu }=j_{\mu}j_{\nu }^*, \ \ H_{\mu\nu }=J_{\mu}J_{\nu }^*.
\label{eq:eq16}
\ee
If the initial and scattered muons are unpolarized, the leptonic
tensor is
\be
L_{\mu\nu }(0)=2q^2g_{\mu\nu}+4(k_{1\mu}k_{2\nu}+k_{2\mu}k_{1\nu}).
\label{eq:eq17}
\ee
where $g_{\mu\nu}$ is the symmetric tensor: $g_{\mu\nu}=(\gamma_\mu\gamma_\nu +\gamma_\nu\gamma_\mu)/2$. In the case of polarized muon beam, the spin-dependent part of
the leptonic tensor can be written as
\be
L_{\mu\nu }(\xi)=2im<\mu\nu q\xi>,
\label{eq:eq18}
\ee
where $<\mu\nu ab>=\varepsilon_{\mu\nu\rho\sigma
}a_{\rho}b_{\sigma}$ and $\xi_{\mu}$ is the muon polarization 4-vector. 

The polarization 4--vector of a particle in the system where its momentum is
${\vec p}$ is connected with the polarization vector ${\vec \chi}$ in its rest
frame by a Lorentz boost $${\vec \xi}={\vec \chi}+\frac{{\vec p}\cdot {\vec
\chi}{\vec p}}{m(E+p)}, \ s^0=\frac{1}{m}{\vec p}\cdot {\vec \xi}. $$
The polarization four vector satisfies the conditions $\xi^2=-1, k_1\cdot \xi=0$.

Let us introduce the following notation:
\be
R_1=|\overline{\cal M}_1|^2,~ R_{int}=2Re\overline{\cal M}_1\overline{\cal M}_2^*.
\label{eq:eq19}
\ee
%%%%%%%%%%%%%%%%%%%%%%%%%%%%%%%%%%%%%
\section{The differential cross section}
%%%%%%%%%%%%%%%%%%%%%%%%%%%%%%%%%%%%%
In this section consider the elastic scattering of unpolarized muon beam by
unpolarized proton target. The hadronic tensor
$H_{\mu\nu}(0)$ can be written as
\be
H_{\mu\nu }(0)=H_1\tilde g_{\mu\nu }+H_2P_{\mu }P_{\nu
}+H_3(P_{\mu}K_{\nu}+P_{\nu}K_{\mu})+iH_4(P_{\mu}K_{\nu}-P_{\nu}K_{\mu}),
\label{eq:eq20}
\ee
with $\tilde g_{\mu\nu }=g_{\mu\nu }-q_{\mu }q_{\nu}/q^2$.
With the hadronic current given by Eq. (\ref{eq:eq4}),
the  structure functions $H_i$, i=1-4, are expressed in terms of the amplitudes as:
\ba
H_1&=&2q^2(G_{M}^2+2G_{M}Re\Delta G_{M}), \nn \\
H_2&=&\frac{8}{1+\tau }\left [G_{E}^2+\tau G_{M}^2+2G_{E}Re\Delta
G_{E}+2\tau G_{M}Re\Delta G_{M}+\right . 
\nn \\
&& 
\left . 2\frac{P\cdot K}{M^2}(\tau
G_{M}+G_E)Re\tilde F_3\right ], \nn \\
H_3&=&-8\tau G_MRe\tilde
F_3, \ \ H_4=-8\tau G_MIm\tilde F_3.
\label{eq:eq21}
\ea
The contraction of the leptonic $L_{\mu\nu}(0)$ and hadronic
$H_{\mu\nu}(0)$ tensors gives
\be
R_1(0)=8m^2(H_1+2P\cdot KH_3)+4q^2H_1+2[(1+\tau )M^2q^2+4(P\cdot
K)^2]H_2.
\label{eq:eq22}
\ee
The interference term $R_{int}$ can be written as
\ba
R_{int}(0)&=&128m^2\left \{ P\cdot KG_ERe
\left [\tilde F_4+\frac{1}{(1+\tau)}\frac{P\cdot K}{M^2}\tilde F_5\right ]+\right .
\nn\\
&&
\left .\tau G_M\left [\frac{1}{(1+\tau
)}\frac{(P\cdot K)^2}{M^2}-K^2\right ]Re\tilde F_5\right \}.
\label{eq:eq23}
\ea
Therefore, the unpolarized differential cross section of the reaction (\ref{eq:eq1}) in
the laboratory (Lab) system, taking into account the lepton
mass and the terms due to the two-photon-exchange,  can be written in the form
\be
\frac{d\sigma_{un}}{d\Omega } = \sigma_0D,
\label{eq:eq24}
\ee
where $\sigma_0 $ is the cross section for the scattering of lepton
on a point-like spin 1/2 particle. It is a generalisation of the Mott
cross section (including a recoil factor) to the case when the lepton mass
is not neglected
\be
\sigma_0=4\frac{\alpha^2}{q^4}\frac{M}{d}\frac{{\vec k}_2^2}{|{\vec
k}_1|}[\varepsilon_1^2-M(M+2\varepsilon_1)\tau ].
\label{eq:eq25}
\ee
Note that in the limit $m=0$ this expression reduces to the Mott
cross section
\be
\sigma_0(m=0)=\frac{\alpha^2\cos^2\displaystyle\frac{\theta}{2}}
{4\varepsilon_1^2\sin^4\displaystyle\frac{\theta}{2}}
\left ( 1+
2\frac{\varepsilon_1}{M}\sin^2\displaystyle\frac{\theta}{2}\right )^{-1},
\label{eq:eq26}
\ee
where $\theta $ is the  Lab muon scattering angle.

The quantity $D$, which contains the information about the structure
of the target and the effects of the two-photon-exchange mechanism, has the form
\ba
D&=&(1+\tau )^{-1}\Bigl [G_E(G_E+2Re\Delta G_E)+\tau G_M(G_M+2Re\Delta G_M)+
\nn\\
&& 
 2\left(\frac{\varepsilon_1}{M}-\tau \right )(G_E+\tau G_M)Re\tilde F_3\Bigr ]+
\nn\\
&& 
\frac{1}{2}\left [\varepsilon_1^2-M(M+2\varepsilon_1)\tau \right ]^{-1}
\Bigl \{-\tau (q^2+2m^2)G_M(G_M+2Re\Delta G_M)- 
\nn\\
&&
4\tau m^2\left(\frac{\varepsilon_1}{M}-\tau  \right )G_MRe\tilde F_3 +
\nn\\
&&
4\tau \frac{m^2}{M^2}(1+\tau )^{-1}\left [\varepsilon_1^2-m^2-\tau
(m^2+M^2)-2\tau M\varepsilon_1 \right ]G_MRe\tilde F_5+ 
\nn\\
&&
\left . 4m^2\left (\frac{\varepsilon_1}{M}-\tau \right )G_E Re\left [\tilde F_4+
(1+\tau )^{-1}(\frac{\varepsilon_1}{M}-\tau )\tilde F_5\right ]\right \}.
\label{eq:eq27}
\ea
In case of one-photon exchange, taking into account the lepton mass, we recover the expressions of Ref. \cite{Preedom:1987mx} and the differential cross section becomes:
\be
\frac{d\sigma}{d(-q^2)}=\frac{\pi\alpha^2}{2M^2\vec k_1^2}\frac{\cal D}{q^4},
\label{eq:eqSk}
\ee
with
\ba
{\cal D}&=&q^2(q^2+2m^2)G_M^2(q^2)+\nn\\
&&2\left [q^2M^2+\frac{1}{1+\tau}\left(2M\epsilon_1+\frac{q^2}{2}\right )^2\right ]\left [ G_E^2(q^2)+\tau G_M^2(q^2)\right ].
\label{eq:eqD2}
\ea
For experiments where the scattered lepton is detected, it may be useful to give the equivalent expression in terms of the solid angle of the lepton,
$ d\Omega=2\pi d\cos\theta $ :
\be
\frac{d\sigma}{d\Omega }=\frac{\alpha^2 |\vec k_2|^3}{2M|\vec k_1|(M\epsilon_1\epsilon_2 -m^2 E_2)}\frac{\cal D}{q^4},
\label{eq:eqSk2}
\ee
where the relation between the energy 
and the angle of the scattered lepton $\theta$ was given in Eq. (\ref{eq:eq12}).

%
%%%%%%%%%%%%%%%%%%%%%%%%%%%%%%%%%%%%%%
\section{Polarization observables}
%%%%%%%%%%%%%%%%%%%%%%%%%%%%%%%%%%%%%%%%%%%%%%%%

The calculation of the polarization observables requires to
define a coordinate frame. Let us specify the coordinate frame in
the Laboratory (Lab) system: the $z$-axis is directed along the muon beam
momentum $\vec{k}_1$, the $y$-axis is directed along the vector
$\vec{k}_1\times \vec{k}_2$, and the $x$-axis is chosen in order to
form a left handed coordinate system. Therefore the reaction plane is the
$xz$-plane.
%%%%%%%%%%%%%%%%%%%%%%%%%%%%%%%%%%%%%%%%
\subsection{T-even polarization observables}
%%%%%%%%%%%%%%%%%%%%%%%%%%%%%%%%%%%%%%%%
In this chapter we consider the Time (T)-even polarization observables,
which depend on the spin correlation $\vec{\xi}\cdot\vec{\xi}_1$
(polarized muon beam and proton target) and $\vec{\xi}\cdot\vec{\xi}_2$
(polarized muon beam and recoil proton). All these T-even
polarization observables are determined by the real parts of the
two--photon--exchange amplitudes.
%%%%%%%%%%%%%%%%%%%%%%%%%%%%%%%%%%%%%%%%
\subsubsection{Polarized beam and target}
%%%%%%%%%%%%%%%%%%%%%%%%%%%%%%%%%%%%%%%%
The hadronic tensor $H_{\mu\nu }(\xi_1)$ corresponding to the
contribution of the non-flip spin amplitudes is
\ba
H_{\mu\nu }(\xi_1)&=&H_5P_{\mu}P_{\nu}+H_6(P_{\mu}<Kq\xi_1\nu
>+P_{\nu}<Kq\xi_1\mu >)+\nn\\
&& 
H_7(P_{\mu}<p_1p_2\xi_1\nu >+
P_{\nu}<p_1p_2\xi_1\mu >)+
\nn\\
&&
iH_8<\mu\nu q\xi_1>+iH_9<\mu\nu qK>+iH_{10}<\mu\nu p_1p_2>,
\label{eq:eq28}
\ea
where $H_i (i=4-10)$ are the structure functions and their
expressions in terms of the amplitudes are
\ba
H_5&=&\frac{4}{M^3}(1+\tau )^{-1}<PKq\xi_1>(G_M-G_E)Im\tilde F_3, ~
H_6=\frac{2}{M}G_MIm\tilde F_3,\nn\\
H_7&=&\frac{2}{M}(1+\tau )^{-1}Im(G_M\Delta G_E-G_E\Delta G_M),\nn\\
H_8&=&-2\left [MG_MG_E+(\varepsilon_1-M\tau )G_MRe\tilde F_3+\right .\nn\\
&&
\left . M(G_MRe\Delta G_E+G_ERe\Delta G_M)\right ], \nn
\\
H_9&=&\frac{1}{M}q\cdot \xi_1G_MRe\tilde F_3,\nn\\
H_{10}&=&\frac{1}{M}(1+\tau )^{-1}q\cdot \xi_1\left (G_M^2-G_MG_E+2G_M\Delta G_M-
\right .
\nn\\
&& \left . G_M\Delta
G_E-G_E\Delta G_M\right ). 
\label{eq:eq29}
\ea
Note that symmetrical part of the tensor
$H_{\mu\nu}(\xi_1)$ is determined by the two-photon-exchange
amplitudes only.

Consider the scattering of the polarized muon beam on the polarized
proton target. In this case we have
\ba
R(\xi, \xi_1)&=&16mM(q\cdot \xi q\cdot \xi_1-q^2 \xi\cdot \xi_1)(G_MG_E+G_MRe\Delta
G_E+\nn\\
&&
G_ERe\Delta G_M)+16mM\frac{\tau}{1+\tau}P\cdot \xi q\cdot \xi_1\left [G_M(G_M-G_E)+ \right .\nn\\
&&
\left . 2G_MRe\Delta
G_M-G_MRe\Delta G_E-G_ERe\Delta G_M\right ]+
\nn\\
&&
8m\left [\varepsilon_1q\cdot \xi q\cdot \xi_1-q^2(\varepsilon_1-M\tau )\xi \cdot
\xi_1\right ]G_MRe\tilde F_3, \nn
\ea
\ba
R_{int}(\xi, \xi_1)&=&8\frac{m}{M}G_M\{-q^2\xi\cdot
\xi_1(K\cdot PRe\tilde F_4+K^2Re\tilde F_5)+
\nn\\ &&
(K^2q\cdot \xi q\cdot
\xi_1+q^2K\cdot \xi K\cdot \xi_1)Re\tilde F_5+
\nn\\ &&
[M\varepsilon_1q\cdot \xi q\cdot
\xi_1+\frac{q^2}{2}(q\cdot \xi q\cdot \xi_1+2p_1\cdot \xi K\cdot \xi_1)]Re\tilde
F_4\}-
\nn\\ &&
8\frac{m}{M}\frac{G_M-G_E}{M^2(1+\tau)}<PKq\xi><PKq\xi_1>Re\tilde
F_5+ \nn\\ 
&&
+2\frac{m}{M}[q^2G_M(q^2\xi\cdot \xi_1-2q\cdot \xi k_1\cdot \xi_1)+
\nn\\ 
&&
2q\cdot \xi_1
\frac{\tau G_M+G_E}{1+\tau}(q^2p_1\cdot \xi-2M\varepsilon_1q\cdot \xi)] Re\tilde F_6.
\ea
The differential cross section of the reaction (\ref{eq:eq1}) describing the
scattering of polarized muon beam on a polarized proton target can
be written as (we give here only spin-dependent part of the cross
section which is determined by the spin correlation coefficients, $C_{ij}$):
\be
\frac{d\sigma(\xi,\xi_1)}{d\Omega}=
\frac{d\sigma_{un}}{d\Omega}(1+C_{xx}\xi_x\xi_{1x}+
C_{yy}\xi_y\xi_{1y}+C_{zz}\xi_z\xi_{1z}+C_{xz}\xi_x\xi_{1z}+C_{zx}\xi_z\xi_{1x}),
\label{eq:eq 26}
\ee
where the
vector $\vec{\xi}(\vec{\xi}_1)$ is the unit polarization vector in
the rest frame of the lepton beam (proton target) and the spin correlation coefficients have the
following form in terms of the amplitudes
%%%%%%%%%%%%%%%%%%%%%%%%%%%
\ba
\overline
DC_{xx}&=&2\frac{m}{M}\left \{ G_M \left [q^2G_E+\frac{\vec{k}_2^2\sin^2\theta}{1+\tau}(G_E+\tau
G_M)\right ]+ \right .\nn\\
&&
q^2(G_ERe\Delta G_M+G_MRe\Delta G_E)+ \nn\\
&&
\left .\frac{\vec{k}_2^2\sin^2\theta}{1+\tau}(G_ERe\Delta G_M+G_MRe\Delta
G_E+2\tau G_MRe\Delta G_M)\right \}+ 
\nn\\
&&
\frac{m}{M}G_M\left \{\left [-\tau
q^2+\frac{\varepsilon_1}{M}(q^2+\vec{k}_2^2sin^2\theta )\right ]Re(2\tilde
F_3+\tilde F_4-\tilde F_6)+\right . \nn\\
&&
\left . \left [\tau
q^2+\frac{m^2}{M^2}(q^2+\vec{k}_2^2\sin^2\theta )\right ]Re\tilde
F_5\right \}+\nn\\
&&
\frac{m\varepsilon_1}{M^2}\left [q^2G_M+\frac{\vec{k}_2^2\sin^2\theta}{1+\tau}
(G_M-G_E)\right ]Re\tilde F_6, 
\label{eq:Cij} 
\ea
%%%%%%%%%%%%%%%%%%%%%%%%%%%
\ba
\overline
DC_{yy}&=&2\frac{m}{M}\Bigl \{ 2q^2(G_MG_E+G_ERe\Delta G_M+G_MRe\Delta
G_E)+ \nn\\
&&
q^2G_M\left [\left (\frac{\varepsilon_1}{M}-\tau \right )Re(2\tilde F_3+\tilde
F_4)+ \left (\tau +\frac{m^2}{M^2}\right ) Re\tilde F_5+\tau Re\tilde F_6\right ]-\nn\\
&&
 \frac{\vec{k}_1^2\vec{k}_2^2\sin^2\theta}{M^2(1+\tau)}(G_M-G_E)Re\tilde
F_5\Bigr \}, \nn
\ea
%%%%%%%%%%%%%%%%%%%%%%%%%%%
\ba
\overline
DC_{xz}&=&-2\frac{m}{M}|\vec{k}_2|(|\vec{k}_1|-|\vec{k}_2|\cos\theta
)\sin\theta \Bigl\{2(1+\tau )^{-1}
\left [G_M(G_E+\tau G_M)+
\right .\nn\\
&&
\left .
2\tau G_MRe\Delta
G_M +G_ERe\Delta G_M+G_MRe\Delta
G_E \right ]+
\nn\\
&&
\frac{\varepsilon_1}{M}G_MRe(2\tilde F_3+\tilde
F_4)+\frac{m^2}{M^2}G_MRe\tilde F_5- \nn\\
&&
\frac{\varepsilon_1}{M}(1+\tau )^{-1}(G_E+\tau G_M)Re\tilde
F_6\Bigr \}+
\nn\\
&&
4\frac{m}{M}\tau |\vec{k}_1||\vec{k}_2|\sin\theta G_MRe(\tilde
F_4+\tilde F_5+\tilde F_6), \nn
\ea
%%%%%%%%%%%%%%%%%%%%%%%%%%%
\ba
\overline
DC_{zx}&=&4\frac{|\vec{k}_2|}{M}\sin\theta \{G_M [2\tau
M|\vec{k}_1|G_M+\varepsilon_1(|\vec{k}_2|\cos\theta - \nn\\
&&
|\vec{k}_1|)(1+\tau )^{-1}(\tau G_M+G_E) ]+ 
4\tau M|\vec{k}_1|G_MRe\Delta
G_M+\nn\\
&&
\varepsilon_1(|\vec{k}_2|\cos\theta -|\vec{k}_1|)
(1+\tau)^{-1}(G_ERe\Delta G_M+G_MRe\Delta G_E+ \nn\\
&&
 2\tau G_MRe\Delta G_M) \}+
\nn\\
&&
2\frac{|\vec{k}_2|}{M}\sin\theta
(\varepsilon_1|\vec{k}_2|\cos\theta
-\varepsilon_2|\vec{k}_1|)G_MRe\left (\frac{m^2}{M^2} \tilde F_5+ 2
\frac{\varepsilon_1}{M} \tilde F_3\right )+\nn\\
&&
2\frac{m^2}{M^2}|\vec{k}_2|
\sin\theta (|\vec{k}_2|\cos\theta -|\vec{k}_1|)\nn\\
&&
\left [G_MRe\tilde
F_4-
(1+\tau )^{-1}(\tau G_M+G_E)Re\tilde F_6\right ], \nn
\ea
%%%%%%%%%%%%%%%%%%%%%%%%%%%
\ba
\overline
DC_{zz}&=&-8\tau [|\vec{k}_1|(|\vec{k}_1|-|\vec{k}_2|\cos\theta
)G_M(G_M+2Re\Delta G_M)+\nn\\
&&
2M\varepsilon_1(G_MG_E+G_ERe\Delta
G_M+G_MRe\Delta G_E)]+ \nn\\
&&
4\frac{\varepsilon_1}{M}(1+\tau )^{-1}(|\vec{k}_1|-|\vec{k}_2|\cos\theta
)^{2}[G_M(\tau G_M+G_E)+G_ERe\Delta G_M+
\nn\\
&&
G_MRe\Delta G_E+2\tau
G_MRe\Delta G_M]- 
2\frac{\varepsilon_1^2}{M^2}\vec{k}_2^2\sin^2\theta G_MRe(2\tilde
F_3+Re\tilde F_4)-
\nn\\
&&
4\tau (\tau M^2-2\vec{k}_1^2)G_MRe\tilde F_4+ 
2\frac{m^2}{M^2}\{2\tau M^2(G_MRe\tilde F_4-
 \nn\\
&&
2\tau G_ERe\tilde
F_6)+\vec{k}_2^2\sin^2\theta [\frac{\tau G_M+G_E}{1+\tau }Re\tilde
F_6-
\frac{\varepsilon_1}{M}G_MRe\tilde F_5]\}, 
\nn
\ea
with $\overline D=[q^2(1+\tau )+4(\varepsilon_1-M\tau )^2]D$.

%%%%%%%%%%%%%%%%%%%%%%%%%%%%%%%%%%
\subsection{T-odd spin observables}
%%%%%%%%%%%%%%%%%%%%%%%%%%%%%%%%%%

In this chapter we consider the T-odd polarization observables which
depend on the T-odd polarization correlations
$\vec{k}_1\times\vec{k}_2\cdot\vec{s}$ (the beam transverse
asymmetry), $\vec{k}_1\times\vec{k}_2\cdot\vec{s}_1$ (the target
normal spin asymmetry), $\vec{k}_1\times\vec{k}_2\cdot\vec{s}_2$
(the normal polarization of the recoil proton). All these T-odd
polarization observables are determined by the imaginary part of
the two-photon-exchange amplitudes.
%%%%%%%%%%%%%%%%%%%%%%%%%%%%%%%%%%
\subsubsection{ Polarized muon beam}
%%%%%%%%%%%%%%%%%%%%%%%%%%%%%%%%%

Let us consider the single spin asymmetry induced by the transverse
polarization of the muon beam. From Eq. (\ref{eq:eq20}) one can see that the 
two-photon-exchange contribution leads to the antisymmetric part
of the spin-independent hadronic tensor $H_{\mu\nu}(0)$. As a result, 
in the general case, a non-zero asymmetry is due to the 
polarization of the muon beam. 

The expressions for the spin-
dependent leptonic tensor and for the spin-independent hadronic
tensor show that the single spin asymmetry is proportional to the
two-photon-exchange term and suppressed by the factor ($m/M$).

The measurement of this small observable has been done and showed that the asymmetry
in the scattering of transversely polarized electrons on unpolarized
protons is different from zero, contrary to what is expected in the
Born (one-photon-exchange) approximation \cite{Maas:2004pd,Wells:2000rx}.

In the case when only the muon beam is polarized, the differential cross section of the reaction (\ref{eq:eq1}) can be written as:
\be
\frac{d\sigma(\xi)}{d\Omega}=\frac{d\sigma_{un}}{d\Omega}(1+A_{\mu}\xi_y),
\label{eq:eq33}
\ee
where $A_{\mu}$ is the beam asymmetry, i.e., the asymmetry 
due to the polarization of the muon beam. Note that
this asymmetry is determined by the $y$-component of the muon
polarization. The expression of the asymmetry in terms of the
amplitudes is
\ba
DA_{\mu}&=&-\frac{m}{M}\frac{|\vec{k}_1||\vec{k}_2|\sin\theta}
{\varepsilon_1^2-M(M+2\varepsilon_1)\tau } Im \Bigl[\tau G_M\tilde
F_3+G_E\tilde F_4+\nn\\
&&
(1+\tau )^{-1}\Bigl (\frac{\varepsilon_1}{M}-\tau \Bigr )(\tau
G_M+G_E)\tilde F_5\Bigr].
\label{eq:eq34}
\ea
Let us enumerate the following properties of this observable:
\begin{itemize}
\item $A_{\mu}$ is proportional to the muon mass and it is determined by
the muon spin component perpendicular to the reaction plane.
\item $A_{\mu}$ is a T-odd observable and it vanishes in the Born
approximation as it is determined by the imaginary part of the
interference between the one- and two-photon exchange amplitudes.
Thus, the asymmetry $A_{\mu}$ is determined by two real
electromagnetic FFs $G_M(q^2)$, and $G_E(q^2)$ as well as by four
complex two-photon exchange induced amplitudes: $\tilde F_3(q^2, s)$
(helicity conserving) and $\tilde F_i(q^2, s)$, $\ (i=4-6)$
(helicity non conserving). Therefore, this observable contains all
amplitudes on equal footing, i.e., here the helicity flip amplitudes
are not suppressed in comparison with the helicity conserving ones.
\item  $A_{\mu}$ vanishes, for $\theta =0^0$ and $180^0$, as it is
determined by the product $\vec{k}_1\times \vec{k}_2\cdot \vec{s}$,
and in this case  $\vec{k}_1|| \vec{k}_2$.
\end{itemize}

%%%%%%%%%%%%%%%%%%%%%%%%%%%%%%%%%%
\subsubsection{Polarized proton target} 
%%%%%%%%%%%%%%%%%%%%%%%%%%%%%%%%%%%%

Let us consider the single spin asymmetry due to the polarized
proton target (called target normal spin asymmetry). Since the
symmetrical part of the spin-dependent hadronic tensor
$H_{\mu\nu}(\xi_1)$ is determined by the two-photon-exchange
amplitudes, the target normal spin asymmetry is also
determined by these amplitudes. The expressions for the
corresponding contractions are the following
\ba 
R_1(\xi_1)&=&\frac{8}{M}<PKq\xi_1>\Bigl \{-q^2G_MIm\tilde F_3+4(1+\tau )^{-1}K\cdot
PIm(G_M\Delta G_E-\nn\\
&&G_E\Delta G_M)
+\Bigl [-q^2+4(1+\tau )^{-1}\frac{(K\cdot P)^2}{M^2}\Bigr ](G_M-G_E)Im\tilde
F_3\Bigr \},
\label{eq:eq35}
\ea
\be
R_{int}(\xi_1)=16\frac{m^2}{M}<PKq\xi_1>[G_MIm\tilde F_4+(1+\tau )^{-1}
\frac{K\cdot P}{M^2}(G_M-G_E)Im\tilde F_5]. 
\label{eq:eq36}
\ee
When only the proton target is polarized, the differential cross section of the reaction (\ref{eq:eq1}) can be written as 
\be
\frac{d\sigma(\xi_1)}{d\Omega}=\frac{d\sigma_{un}}{d\Omega}(1+A_{p}\xi_{1y}),
\label{eq:eq37}
\ee
where $A_{p}$ is the target asymmetry, i.e., the asymmetry due to the 
polarization of the proton target.
Note that this asymmetry is determined by the $y$-component of the
proton polarization vector. From Eq. (\ref{eq:eq36}) one can see that the
contribution of the spin flip amplitudes is suppressed by the factor
$m^2/M^2$ which is small also in the case of the muon scattering. So, neglecting 
this contribution,  the asymmetry $A_p$ can be written in terms of the amplitudes as
\ba
DA_{p}&=&-\frac{|\vec{k}_1||\vec{k}_2|\sin\theta}
{\varepsilon_1^2-M(M+2\varepsilon_1)\tau }
\left \{\tau G_EIm\tilde F_3+(1+\tau )^{-1}\frac{K\cdot P}{M^2}
\right .
\nn\\
&&
\left . \left [G_MIm\Delta G_E-G_EIm\Delta G_M+\frac{K\cdot P}{M^2}(G_M-G_E)Im\tilde F_3\right ]\right \}. 
\label{eq:eq38}
\ea
One can see that
\begin{itemize}
\item $A_p$ is determined by the component of the proton polarization
vector perpendicular to the reaction plane, i.e., by the following
product $\vec{k}_1\times\vec{k}_2\cdot\vec{s}_1$;
\item $A_p$ vanishes when $\vec{k}_1||\vec{k}_2$, i.e., in collinear
kinematics;
\item $A_p$ vanishes in the Born approximation. It is determined by the
interference of the one- and two-photon-exchange amplitudes, through the
imaginary parts of all three complex two-photon-exchange helicity
conserving amplitudes.
\end{itemize}
%%%%%%%%%%%%%%%%%%%%%%%%%%%%%%%%%%%%%%%%%%%%%%%%%%%%%%%%%%%%%%%%
\subsubsection{Polarization of the recoil proton}
%%%%%%%%%%%%%%%%%%%%%%%%%%%%%%%%%%%%%%%%%%%%%%%%%%%%%%%%%%%%%%%%

Let us consider the polarization of the recoil proton in the case when all
the other particles are not polarized. In the Born approximation such
polarization vanishes. The interference between the one- and the two-photon-exchange contributions induce a non-zero polarization of the recoil
proton. Note that only a measurement exists for such single spin observables: the
recoil-deuteron vector polarization in unpolarized electron deuteron
elastic scattering \cite{PhysRevLett.21.1271}.

The spin-dependent part of the hadronic tensor $H_{\mu\nu }(\xi_2)$
which describes the case of the polarized recoil proton can be
obtained from Eq. (\ref{eq:eq28}) by the following substitution: $\xi_1\to \xi_2$. The structure
functions describing this tensor can be obtained in the same way. In
this case the symmetric part of the tensor $H_{\mu\nu }(\xi_2)$ is
also determined by the interference of the one- and
two-photon-exchange contributions. The contractions $R_1$ and
$R_{int}$ in this case can be written as
\ba
R_1(\xi_2)&=&32M<PKq\xi_2>\Bigl \{\tau G_MIm\tilde F_3+(1+\tau
)^{-1} \left [\frac{\varepsilon_1^2}{M^2}-\left (1+2\frac{\varepsilon_1}{M}\right ) \right ] 
\nn\\
&&
(G_M-G_E)Im\tilde F_3+m(1+\tau
)^{-1}\left (\frac{\varepsilon_1}{M}-\tau \right )\nn\\
&&
(G_M\Delta G_E-G_E\Delta
G_M)\Bigr \}, \nn\\
R_{int}(\xi_2)&=&32\frac{m^2}{M}<PKq\xi_2>\left [G_MIm\tilde F_4+(1+\tau
)^{-1} \left (\frac{\varepsilon_1}{M}-\tau \right )\right .\nn\\
&&
\left .
(G_M-G_E)Im\tilde F_5\right ].
\label{eq:eq42}
\ea
The differential cross section of the reaction (\ref{eq:eq1}) can be written, 
for the case when only the recoil proton is polarized, as:
\begin{equation}\label{23}
\frac{d\sigma(\xi_2)}{d\Omega}=\frac{d\sigma_{un}}{d\Omega}\left (1+P_y\xi_{2y}\right ),
\end{equation}
where $P_{y}$ is the $y-$ component of the recoil-proton
polarization vector and $\vec{\xi_2}$ is the unit polarization
vector in the rest frame of the recoil proton. Note that this
polarization is determined by the $y$-component of the recoil-proton
polarization vector. From this expression one can see that the
contribution of the spin flip amplitudes is suppressed by the factor
$m^2/M^2$ which is small also in the case of the muon scattering.
So, neglecting this contribution, the polarization $P_y$ can be
written in terms of the amplitudes as
\ba
DP_{y}&=&-\frac{|\vec{k}_1||\vec{k}_2|\sin\theta}
{\varepsilon_1^2-M(M+2\varepsilon_1)\tau } \Bigl \{(1+\tau )^{-1}\left(\frac{\varepsilon_1}{M}-\tau \right )\Bigl [G_MIm\Delta G_E-G_EIm\Delta G_M+
\nn\\
&&
+\left (\frac{\varepsilon_1}{M}-\tau \right ) (G_M-G_E)Im\tilde F_3\Bigr ]+\tau
G_EIm\tilde F_3 \Bigr \}. 
\label{23a}
\ea
One can see that
\begin{itemize}
\item $P_y$ is determined by the component of the proton polarization
vector perpendicular to the reaction plane, i.e., by the following
product $\vec{k}_1\times\vec{k}_2\cdot\vec{\xi}_2$;
\item $P_y$ vanishes when $\vec{k}_1||\vec{k}_2$, i.e., in collinear
kinematics;
\item $P_y$ vanishes in the Born approximation. It is determined by the
interference of the one- and two-photon-exchange amplitudes, through
the imaginary parts of all three complex two-photon-exchange
helicity conserving amplitudes.
\end{itemize}

%%%%%%%%%%%%%%%%%%%%%%%%%%%%%%%%%%%%%%%%%%%%%%%%%%%%%%%%%%%
\subsubsection{Polarized beam and polarized recoil proton}
%%%%%%%%%%%%%%%%%%%%%%%%%%%%%%%%%%%%%%%%%%%%%%%%%%%%%%%%%%%%

In this section, we consider the polarization observables when the 
muon beam and the recoil proton are polarized (the polarization of the recoil proton
is measured).

In this case, the hadronic tensor $H_{\mu\nu }(\xi_2)$ describing the contribution
of the helicity conserving amplitudes has the following form:
\ba
H_{\mu\nu }(\xi_2)&=&g_1P_{\mu}P_{\nu}+g_2(P_{\mu}
<Kq\xi_2\nu>+P_{\nu}<Kq\xi_2\mu >)+\nn\\
&&
g_3(P_{\mu}<p_1p_2\xi_2\nu>+ P_{\nu}<p_1p_2\xi_2\mu >)+ig_4<\mu\nu qK>+\nn\\
&&
ig_5<\mu\nu q\xi_2>+ig_{6}<\mu\nu p_1p_2>, 
\label{eq:eq41}
\ea
where $g_i (i=1-6)$ are the structure functions. Their
expressions in terms of the amplitudes are
\ba
&g_1=&\frac{4}{M^3}(1+\tau )^{-1}<PKq\xi_2>(G_M-G_E)Im\tilde F_3,~
g_2=\frac{2}{M}G_MIm\tilde F_3, \label{eq:eq42b}\\
&g_3=&\frac{2}{M}(1+\tau )^{-1}Im(G_M\Delta G_E-G_E\Delta G_M), ~~~~~~~~
g_4=-\frac{1}{M}q\cdot \xi_2G_MRe\tilde F_3,\nn\\
&g_5=&-2M[G_MG_E+(\frac{\varepsilon_1}{M}-\tau )G_MRe\tilde F_3+
G_MRe\Delta G_E+G_ERe\Delta G_M],\nn\\
&g_{6}=&\frac{1}{M}(1+\tau )^{-1}q\cdot \xi_2[(G_E-G_M)(G_M+Re\Delta G_M)+G_MRe(\Delta
G_E-\Delta G_M)].
\nn
\ea
Note that the symmetrical part of the tensor
$H_{\mu\nu}(\xi_2)$ is determined by the interference of the two- and
one-photon-exchange amplitudes only.

In this case we have
\ba
R(\xi, \xi_2)&=&16mM(q\cdot \xi q\cdot \xi_2-q^2 \xi\cdot \xi_2)(G_MG_E+G_MRe\Delta
G_E+G_ERe\Delta G_M)+\nn\\
&&
32m M\frac{\tau}{1+\tau}P\cdot \xi q\cdot \xi_2[(G_E-G_M)(G_M+Re\Delta
G_M)+\nn\\
&&
G_MRe(\Delta G_E-\Delta G_M)]+ 16\frac{m}{M}
\left [\left (p_1\cdot k_1-2M^2\tau )q\cdot sq\cdot \xi_2-
\right. \right .\nn\\
&&
\left . \left .q^2(p_1\cdot k_1-M^2\tau \right)s\cdot \xi_2 \right]G_MRe\tilde F_3,
\label{eq:eq42a}
\ea
\ba
R_{int}(\xi, \xi_2)&=&8\frac{m}{M}G_M[-q^2\xi\cdot
\xi_2(K\cdot PRe\tilde F_4+K^2Re\tilde F_5-M^2\tau Re\tilde
F_6)+
\nn\\
&&
q\cdot sq\cdot \xi_2(K\cdot PRe\tilde F_4+K^2Re\tilde
F_5-p_1\cdot k_1Re\tilde F_6)+ \nn\\
&&
q^2K\cdot sK\cdot \xi_2Re(\tilde F_5-\tilde F_6)+q^2P\cdot s(K\cdot \xi_2Re\tilde
F_4-P\cdot \xi_2Re\tilde F_6)]\nn\\
&&
+8\frac{m}{M}\frac{G_M-G_E}{1+\tau }[P\cdot \xi_2(q^2P\cdot s-2K\cdot
Pq\cdot s)Re\tilde F_6-
\nn\\
&&
\frac{1}{M^2}<PKq\xi ><PKq\xi_2>Re\tilde F_5].
\label{eq:eq43}
\ea
The dependence of the differential cross section of the reaction (\ref{eq:eq1})
on the polarization transfer coefficients $T_{ij}$, $(i,j=x,y,z)$, 
describing the
scattering of polarized muon beam on an unpolarized proton target
and measuring the recoil-proton polarization,  can be written as
\be
\frac{d\sigma(\xi,\xi_2)}{d\Omega}=\frac{d\sigma_{un}}{d\Omega}(1+T_{xx}\xi_x\xi_{1x}+
T_{yy}\xi_y\xi_{1y}+T_{zz}\xi_z\xi_{1z}+T_{xz}\xi_x\xi_{1z}+T_{zx}\xi_z\xi_{1x}),
\label{eq:eq44}
\ee
The polarization transfer coefficients have the following form in terms
of the FFs and two-photon-exchange amplitudes: 
\ba
%%%%%%%%%%%%%%%%%%%%%%%%%%%%
\bar DT_{xx}&=&2\frac{m}{M}G_M\left [q^2G_E+(1+\tau )^{-1}k_2^2\sin^2\theta
(G_E\right .
\nn\\
&&
\left .-\tau G_M)\right ]+ 2\frac{m}{M}\Bigl \{ q^2\left (G_MRe\Delta G_E+G_ERe\Delta G_M\right )+
\nn\\
&&
G_MRe\tilde F_3
\left(\frac{\varepsilon_1}{M}k_2^2\sin^2\theta -\tau q^2-4\tau M\varepsilon_1\right )+
\nn\\
&&
(1+\tau )^{-1}k_2^2\sin^2\theta
\Bigl[G_MRe(\Delta G_E-2\tau \Delta G_M)+G_ERe\Delta G_M-\nn\\
&&
\left . \left . 2\tau \left (1+\frac{\varepsilon_1}{M}\right )G_MRe\tilde
F_3\right ]\right \}+2\frac{m}{M}q^2G_M\left [\left (\frac{\varepsilon_1}{M}-
\tau \right )Re\tilde F_4+ \right .
\nn\\
&&
\left .\left (\frac{m^2}{M^2}+\tau \right )Re\tilde F_5-\tau Re\tilde
F_6\right ]-2\frac{m}{M}\tau k_2^2\sin^2\theta \Bigl \{G_MRe(\tilde F_5-2\tilde
F_4)+  
\nn\\
&&
2(1+\tau )^{-1}\Bigl[ G_ERe\tilde
F_6+\left (\frac{m^2}{M^2}+\frac{\varepsilon_1}{M}\right )G_MRe\tilde
F_5-
\nn\\
&&
\frac{1}{2M}(\varepsilon_1+\varepsilon_2)G_MRe\tilde F_6 \Bigr \}, 
\nn\\
%%%%%%%%%%%%%%%%%%%%%%%%%%%%
\bar DT_{yy}&=&2\frac{m}{M}\Bigl \{q^2\left(G_MG_E+G_MRe\Delta G_E+G_ERe\Delta
G_M\right )+\nn\\
&&
q^2G_M\left[ \left (\frac{\varepsilon_1}{M}-\tau \right )(2Re\tilde F_3+Re\tilde
F_4)+
\left (\frac{m^2}{M^2}+\tau \right )Re\tilde F_5-\tau Re\tilde
F_6\right ]-\nn\\
&&
 \frac{1}{M^2}k_1^2k_2^2\sin^2\theta (1+\tau
)^{-1}(G_M-G_E)Re\tilde F_5 \Bigr \}, \nn\\
%%%%%%%%%%%%%%%%%%%%%%%%%%%%%%%
\bar DT_{xz}&=&2\frac{m}{M}(1+\tau )^{-1}k_2(k_1-k_2\cos\theta)
\sin\theta \Bigl [(\tau G_M-G_E)(G_M-Re\Delta G_M)+ 
\nn\\
&&
\left .G_M(\tau Re\Delta G_M-
Re\Delta G_E)-\frac{1}{M}(\varepsilon_1-\tau
\varepsilon_1-2\tau M)G_MRe\tilde
F_3\right ]-
\nn\\
&&
2\frac{m}{M}\frac{k_2}{M}\sin\theta
G_M \left [\frac{1}{M}(k_1-k_2\cos\theta )Re(M\varepsilon_1\tilde F_4+ 
m^2\tilde F_5)-\right .
\nn\\ &&
2\tau Mk_2\cos\theta Re(\tilde F_6+\tilde F_4-\tilde
F_5)\Bigr ]+
\nn\\
&&
2\frac{m}{M}\frac{\varepsilon_1}{M}(1+\tau
)^{-1}k_2(k_1-k_2\cos\theta )\sin\theta (G_E+ \tau G_M)Re\tilde F_6-
\nn\\
&&
2\frac{m}{M}(1+\tau
)^{-1}\frac{k_2}{M}\left (k_1-k_2\cos\theta \right )\sin\theta G_M
\Bigl \{\tau
(\varepsilon_1+\varepsilon_2)Re(\tilde F_6+\tilde F_4-\tilde F_5)+
\nn\\&&
\left . (\varepsilon_2-\varepsilon_1)\left [\left(1+\frac{\varepsilon_1}{M}\right)Re\tilde
F_4+\left (\frac{m^2}{M^2}-1 \right ) Re\tilde F_5+Re\tilde F_6\right]\right\}, 
\nn\\
%%%%%%%%%%%%%%%%%%%%%%%
\bar DT_{zx}&=&2\frac{k_1k_2}{M}(1+\tau )^{-1}\sin\theta
\{[k_1-k_2\cos\theta -2(1+\tau )M]\tau G_M(G_M+2Re\Delta G_M)+
\nn\\
&&
(k_2\cos\theta -\varepsilon_1)(G_MG_E+G_MRe\Delta G_E+G_ERe\Delta G_M)-
\nn\\
&&
\frac{1}{M}G_MRe\tilde F_3[(1+\tau
)\varepsilon_1(\varepsilon_1-k_2\cos\theta )+ 
\nn\\
&&
2\tau (M\tau k_1+Mk_2\cos\theta-
k_1\varepsilon_1)]\}+
\nn\\
&&
4\tau (1+\tau
)^{-1}m^2\frac{k_2}{k_1}\frac{\varepsilon_1+\varepsilon_2}{M}G_M\sin\theta
\left [\left (\tau -
\frac{\varepsilon_1}{M}\right )Re\tilde F_5- 
\tau Re\tilde F_6-(1+\tau )Re\tilde F_4\right ]-
\nn\\
&&
4\tau m^2\frac{k_2}{k_1}\sin\theta
\left [\frac{\varepsilon_2}{M}G_MRe(\tilde F_5- \tilde F_4-\tilde
F_6)+ G_MRe \left (\tilde F_4-\tilde F_6
-\frac{m^2}{M^2}\tilde F_5\right )
\right .
\nn\\
&&
\left .
+\left (1+\frac{\varepsilon_1}{M}\right )(1+\tau
)^{-1}(G_E-G_M)Re\tilde F_6\right ], \nn\\
%%%%%%
\bar DT_{zz}&=&2(1+\tau )^{-1}\Bigl \{2\tau^2(M+\varepsilon_1)[2M(1+\tau )+k_2\cos\theta -k_1]G_M
\nn\\
&&
(G_M+2Re\Delta G_M)+
\frac{1}{M}\left [2M\tau
(M+\varepsilon_1)(\varepsilon_1-k_1)-k_1k_2^2\sin^2\theta
\right ]
\nn\\
&&
(G_MG_E+G_MRe\Delta G_E+ 
G_ERe\Delta G_M)\Bigr \}+
\nn\\
&&
\frac{2}{M^2}(1+\tau )^{-1}G_MRe\tilde F_3\{k_1^2k_2^2(\tau\varepsilon_2-
\varepsilon_1)\sin^2\theta +
\nn\\
&&
2M\tau (1+\tau
)[\varepsilon^2_1(M+\varepsilon_1-k_1)- 
Mk_1(1+2\tau )(\varepsilon_1-2\tau M)]\}-
\nn\\
&&
8m^2\tau^2\frac{(M+\varepsilon_1)^2}{k_1^2}
\frac{\tau G_M+G_E}{1+\tau }Re\tilde
F_6+\nn\\
&&
8\tau^2\frac{m^2}{k_1^2}G_MRe(\tilde F_6+ 
\tilde F_4-\tilde F_5)\Bigl [\frac{k_1^2k_2^2}{2M^2(1+\tau
)}\sin^2\theta + 
\nn\\
&&
(1+2\tau
)(M\varepsilon_1+m^2)-
\varepsilon_2(\varepsilon_1+M)\Bigr ]+ \nn\\
&&
8\tau \frac{m^2}{M}G_MRe\tilde F_5\left [\varepsilon_2-\frac{\varepsilon_1k_2^2}
{2M^2(1+\tau )}\sin^2\theta+\right .
\nn\\
&&
 \left . \tau
(M\varepsilon_1+m^2)\frac{M+\varepsilon_1}{k_1^2}\right ]- \nn\\
&&
8\tau G_MRe\tilde F_4 \left \{m^2\left [1-\tau \frac{(M+\varepsilon_1)^2}{k_1^2}\right ]-
\frac{\varepsilon^2_1k_2^2}{2M^2(1+\tau )}\sin^2\theta +\right.
\nn\\
&&
2\tau
\left . \left [\varepsilon^2_1-\tau \frac{(M+\varepsilon_1)^2}{1+\tau }\right ]\right\}. 
\label{eq:Tij}
\ea

%%%%%%%%%%%%%%%%%%%%%%%%%%%%%%%%%%%%%%%%%%%%%
\section{Numerical applications}
%%%%%%%%%%%%%%%%%%%%%%%%%%%%%%%%%%%%%%%%
In this section we illustrate the results through numerical applications. The explicit consideration of the lepton mass modifies the kinematical variables as well as the observables. The reaction under consideration, being a binary process, two variables define completely the kinematics. Therefore, the results are preferentially illustrated as bi-dimensional plots as function of the muon beam energy and the muon scattering angle.
 
\begin{figure}[h]
\begin{center}
\mbox{\epsfxsize=10.cm\leavevmode \epsffile{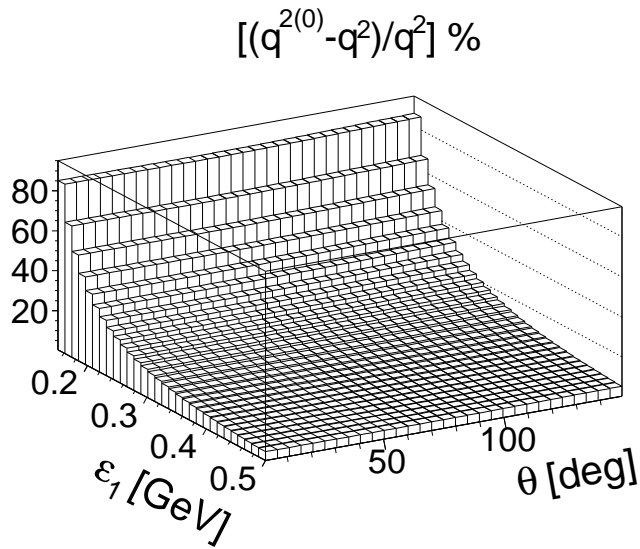}}
%\vspace*{.2 truecm}
\caption{Bi-dimensional plot of the relative difference of momentum transfer squared as function of the incident energy and of the muon scattering angle.} 
\label{Fig:qq}
\end{center}
\end{figure}

The effect of the muon mass is already visible on the momentum transfer squared. The relative difference of the momentum transfer squared $q^2$ taking and not taking into account the lepton mass is shown in Fig. \ref{Fig:qq} as a bi-dimensional plot as function of the beam energy $\epsilon_1$ and the muon scattering angle $\theta$, in the relevant kinematical domain. 
The momentum transfer squared
$$q^2=2m^2-2(\epsilon_1\epsilon_2-k_1k_2\cos\theta)$$
is larger for electron than for a muon at the same incident energy and scattering angle. This difference is comparatively larger at small beam energies.

In order to calculate the unpolarized and polarized observables, the amplitudes have to be calculated according to a model.

In the one-photon exchange approximation, the amplitudes reduce to the two electric $G_E$ and magnetic $G_M$ FFs, which we parametrize for simplicity according to a dipole dependence:
\ba
G_E(Q^2)=G_M(Q^2)/\mu_p=[1+Q^2/0.71]^{-2},
\label{eq:ffs}
\ea
where $\mu_p$ is the anomalous magnetic moment of the proton and $Q^2$ is expressed in [GeV$^2$] units. This parametrization is reasonable in the low $Q^2$ range considered here.
 
In the presence of two-photon exchange, the amplitudes are model dependent. The outcome of models can not be validated on experimental data as no clear evidence of two-photon exchange has emerged from the data up to now. In order not to mislead the reader, we will illustrate the effect of the mass on the observables limiting the calculation to the one-photon exchange approximation. We stress, however that the complete expressions derived in this work are model independent and applicable to any model chosen for the amplitudes.

\begin{figure}[h]
\begin{center}
\mbox{\epsfxsize=14.cm\leavevmode \epsffile{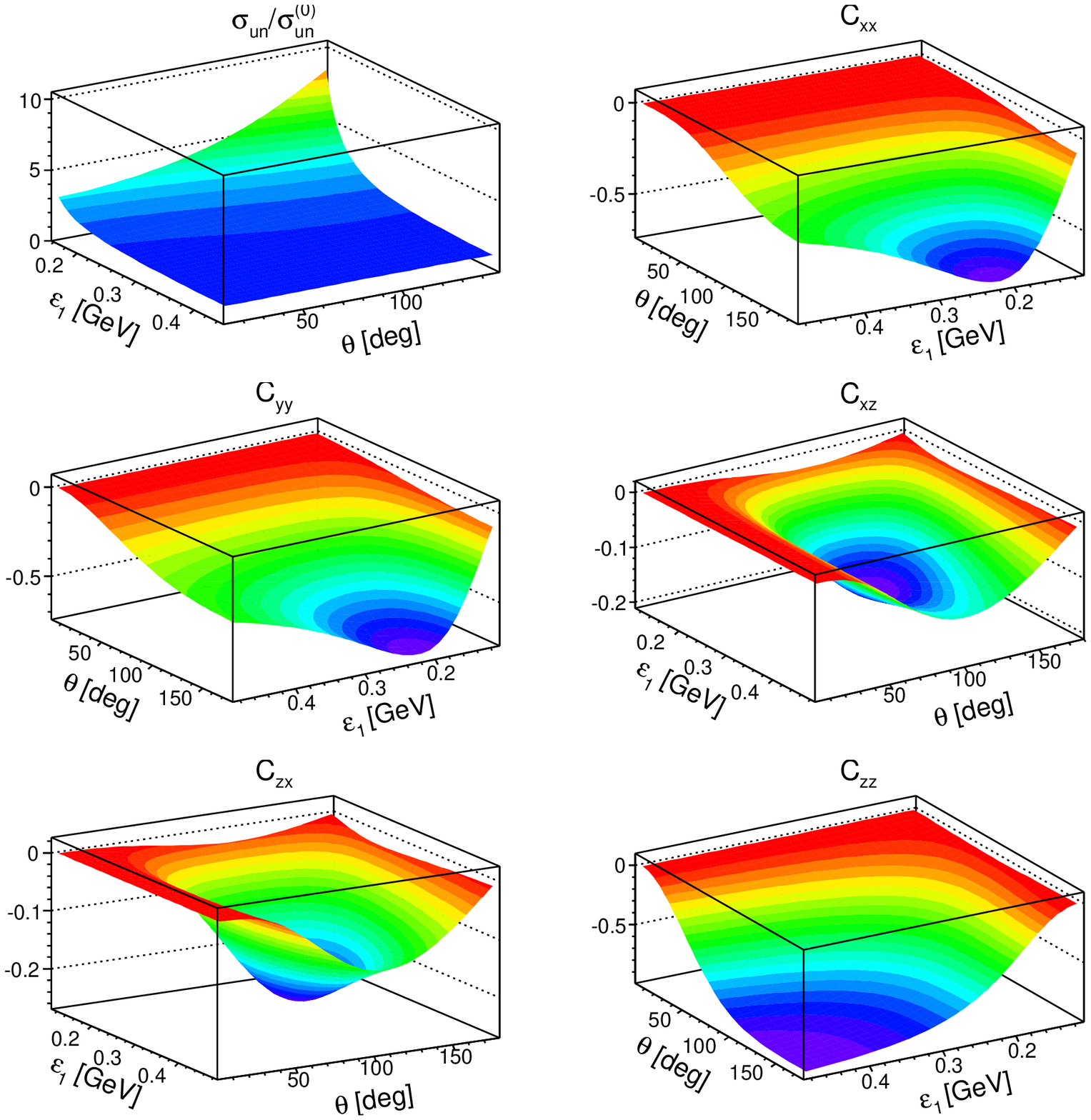}}
%\vspace*{.2 truecm}
\caption{Observables as functions of the beam energy and of the lepton scattering angle. From left to right, from top to bottom: 
ratio of the muon over electron unpolarized elastic scattering cross section  and spin correlation coefficients: 
$C_{xx}$, $C_{yy}$, $C_{xz}$, $C_{zx}$, and $C_{zz}$. 
}
\label{Fig:Cobser}
\end{center}
\end{figure}

The ratio between the cross section taking and not taking into account the lepton mass is shown as function of $\varepsilon_1$ and $\theta$ is shown in Fig \ref{Fig:Cobser}. One can see that this ratio increases essentially at small energies and large angles. This quantity is larger for muons than for electrons, particularly at low energies. As for electrons, the Born cross section diverges for small values of the transferred momentum, i.e., at small incident energy and/or small scattering angles.

The spin correlation coefficients when the initial particles are polarized (from Eqs. (\ref{eq:Cij})) are illustrated in Fig.  \ref{Fig:Cobser}. They are sizable in particular at large angles,  and show a characteristic behavior with energy and angle. The coefficients $C_{xx}$, $C_{yy}$, and $C_{xz}$ are proportional to the lepton mass. The  lepton mass for  $C_{zx}$, and $C_{zz}$ enters explicitly at the level of the two photon amplitudes, enhancing their effect for massive leptons.

\begin{figure}[h]
\begin{center}
\mbox{\epsfxsize=14.cm\leavevmode \epsffile{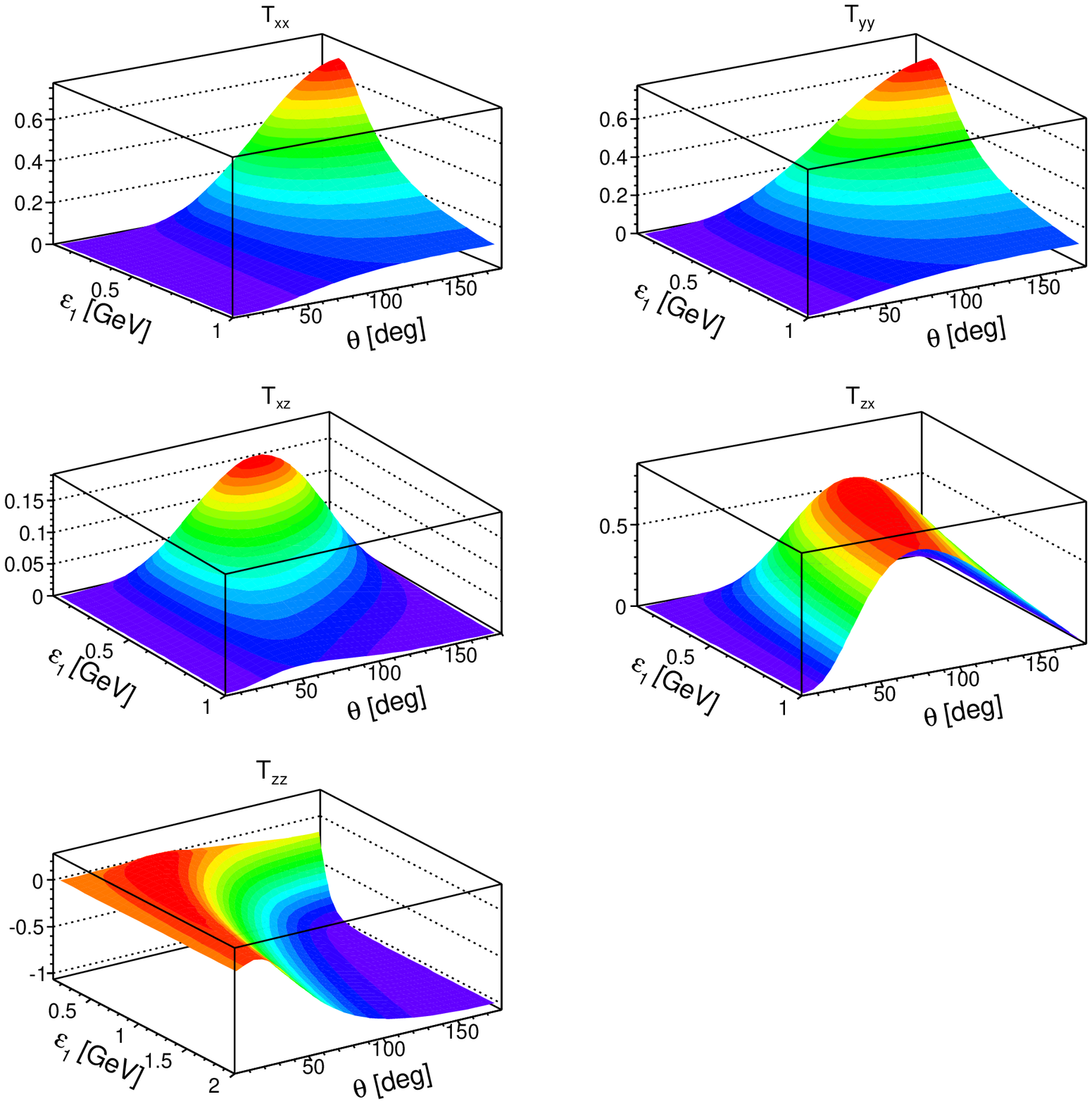}}
%\vspace*{.2 truecm}
\caption{Spin transfer coefficients as functions of the beam energy and of the lepton scattering angle. From left to right, from top to bottom: 
$T_{xx}$, $T_{yy}$, $T_{xz}$, $T_{zx}$, and $T_{zz}$. 
}
\label{Fig:Tobser}
\end{center}
\end{figure}

The spin transfer coefficients, when the beam is polarized and the polarization of the recoil proton is measured, (from Eqs. (\ref{eq:Tij})), are illustrated in Fig.  \ref{Fig:Tobser}. 

The coefficients $T_{xx}$, $T_{yy}$, $T_{xz}$ are proportional to the lepton mass. The mass here also plays a large role in the terms corresponding to the two-photon amplitudes. As for the spin correlations, the spin transfer coefficients show a characteristic behavior, as function of the considered kinematical variables.
%%%%%%%%%%%%%%%%%%%%%%%%
\section{Conclusions}
%%%%%%%%%%%%%%%%%%%%%%%%
The lepton hadron elastic interaction has revisited. The cross section and various polarization observables have been calculated taking into account the lepton mass, and the possible presence of two photon exchange.

The matrix element has been parametrized in the most general form, by six complex amplitudes. Model independent expressions of the observables have been given as functions of these amplitudes.

These expressions are directly applicable to low energy muon proton scattering experiments. Numerical applications have been done in the energy domain covered by planned experiments, in the one photon approximation, for dipole parametrization of the electromagnetic FFs. 

The spin correlation coefficients (when the muon beam and the proton target are polarized) and the spin transfer coefficients (when the muon beam is polarized and the polarization of the recoil proton is measured) have been calculated in the relevant domain of beam energy and muon scattering angle.

It is shown that, in general, polarization observables are sizable and manifest a characteristic angular and energy dependence.
%%%%%%%%%%%%%%%%%%%%%%%%
\section{Acknowledgments}
%%%%%%%%%%%%%%%%%%%%%%%%

This work was partly supported by the French-Ukrainian agreement PICS-5419. Fruitful discussions in frame of French GDR-PH-QCD are acknowledged. Thanks are due to Yu. Bystritskiy for interest in this work and useful advices.

%\bibliographystyle{elsarticle-num}
%\bibliography{Biblio.bib}

\end{document}